# Continuous-Variable Quantum Key Distribution over 15 km Multi-Core Fiber

S. SARMIENTO,[1,6] S. ETCHEVERRY,[1,2,6] J. ALDAMA,[1,6] I. H. LÓPEZ,[1] L. T. VIDARTE,[1] G. B. XAVIER,[3] D. A. NOLAN,[4] J. S. STONE,[4] M. J. LI,[4] D. LOEBER,[4] AND V. PRUNERI[1,5,7]

[1]*ICFO-Institut de Ciències Fotòniques, The Barcelona Institute of Science and Technology, Castelldefels, Barcelona 08860, Spain*
[2]*Department of Physics, Stockholm University, AlbaNova University Center, Stockholm, Sweden*
[3]*Institutionen för Systemteknik, Linköpings Universitet, Linköping 581 83, Sweden.*
[4]*Corning Incorporated, Corning, NY 14831, USA*
[5]*ICREA-Institució Catalana de Recerca i Estudis Avançats, 08010 Barcelona, Spain*
[6]*These authors contributed equally to this work*
[7]*valerio.pruneri@icfo.eu*

**Abstract:** The secure transfer of information is critical for the ever-increasing demands of the digital world. Continuous-variable quantum key distribution (CV-QKD) is a technology that can potentially provide information-theoretic security to cryptographic systems. CV-QKD is highly attractive thanks to high secret key rate generation over metropolitan distances and integration with the current classical communication networks. Here we demonstrate CV-QKD across a 15 km multi-core fiber (MCF) where we take advantage of one core to remotely frequency lock Bob's local oscillator with Alice's transmitter. We also demonstrate the capacity of the MCF to boost the secret key rate by parallelizing CV-QKD across multiple cores. Our results point out that MCFs are promising for metropolitan deployment of QKD systems.

## 1. Introduction

Quantum key distribution (QKD) is an innovative technology used to securely distribute cryptographic keys among spatially-separated users [1, 2]. It is based on encoding randomly chosen bits on individual quantum states, and then performing independent measurements on them. Together with classical post-processing techniques aided by a classical communication channel, a secure and shared secret key can be distilled by the remote parties (typically referred to as Alice and Bob). Over the years, many experiments have been performed showing that QKD is now a mature technology [3-7]. QKD protocols can be divided into two broad categories: discrete-variable (DV) and continuous-variable (CV) QKD [1, 2]. In the former, a discrete set of quantum states are employed together with single-photon detectors [1, 2], while in the later a broader set of states together with coherent detection is employed [8]. CV-QKD has recently gained significant attention as it can be implemented with conventional telecom components, leading to cost-effective implementations that are compatible with current network infrastructures.

As it is widely known, multi-core fibers (MCFs) will play a fundamental role in future classical communications for various reasons. Firstly, MCFs will solve the incoming network capacity crunch. According to [9], the achievable rate of an N-core MCF corresponds theoretically to N times the achievable rate of a single-mode fiber (SMF). Secondly, MCFs present a lower footprint compared a bundle of standard SMFs, which is of high importance in the deployment process, e.g., in telecom data-centers, where space is limited, and they allow the use of a single amplifier for all the cores, reducing the number of resources and increasing energy efficiency [10]. Thirdly, MCFs shows a perfect match with photonic integrated circuits for multiple-input-multiple-output applications and are expected to be widely adopted in long/mid-haul connections [11]. Finally, MCFs have shown similar losses to standard SMFs and small crosstalk between the different cores making them a good solution for co-propagating quantum and classical signals in different cores or in the same core [12-15].

MCFs also show benefits for quantum communication, e.g., the greater phase stability among the different cores due to the fact that co-propagated signals in different cores experience similar environmental effects [16]. MCFs are also well suited to the transmission of high-dimensional photonic states that use the transverse degree-of-freedom of a single-photon [17]. For example, MCFs were initially used to perform QKD sessions with path-encoded qudits over a distance of 300 m [6], and with the use of silicon integrated photonics circuits [18]. Then the distance employed for quantum communication protocols was extended in other experiments to 2 and 11 km, respectively [16, 19]. MCFs have also been used to construct multi-port beam splitters for applications in quantum information [20].

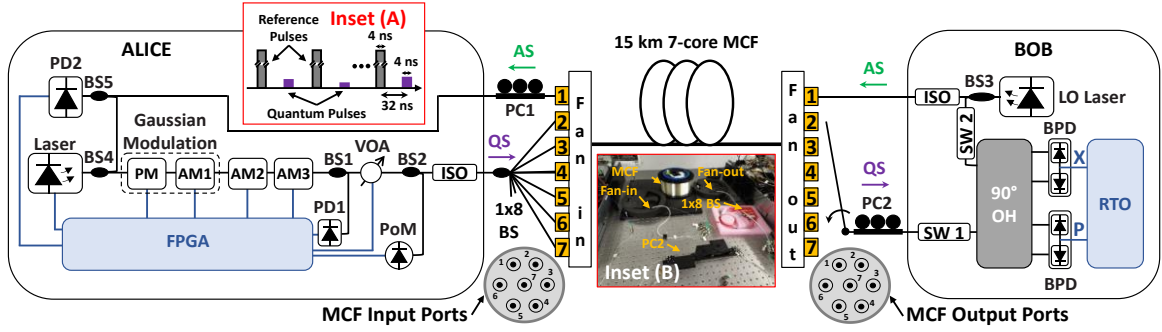

**Fig. 1.** Experimental setup. Inset (A) is the transmitted signal from Alice to Bob, where reference symbols are interleaved between the quantum symbols while Inset (B) is the laboratory setup for the 15 km 7-core MCF spool under test. (Acronyms) FPGA: field-programmable gate array, PM: phase modulator, AM: amplitude modulator, PD: photodiode, BS: fiber beam splitter, VOA: variable optical attenuator, ISO: optical isolator, PC: polarization controller, PoM: power meter, MCF: multi-core fiber, AS: auxiliary signal used to lock the frequency of Alice's laser to the LO, QS: quantum signal, LO: local oscillator, SW: optical switch, 90° OH: 90° optical hybrid, BPD: balanced photodetector, RTO: real-time oscilloscope.

CV-QKD on MCFs have been little studied. In 2018, it was demonstrated that the fan-in and -out devises required in MCFs degrade the performance of CV-QKD protocols [21], but the total rate can be increased through the parallelization. In that same year, a proof-of-concept of a quantum-to-the-home network using CV-QKD through a 9.8 km 7-core MCF was shown [22]. In [22], the authors stated that an aggregated secret key rate of 33.6 Mbps was obtained by using a discrete modulation with four symbols. Nevertheless, no information is provided on how the secret key is calculated nor specific details of the implementation are shown. In 2019, a 10.1 km 19-core MCF was used to demonstrate classical and QKD transmission [15]. In this case, 6 cores were used for QKD and 13 cores for classical coherent signals. 46 Mbps per core and wavelength was achieved. In this work, we implement a Gaussian-modulated coherent-state (GMCS) CV-QKD protocol [23] with a true local oscillator (LO) [24-27] across a 15 km MCF, the longest distance yet for CV-QKD over these fibers. We show that GMCS CV-QKD protocol can be deployed in MCF to boost the secret key rate by parallelizing CV-QKD across multiple cores. In particular, a potential aggregated secret key rate value of 2.3 Mb/s is achieved. In addition, we show that one of the cores can be used to transmit an auxiliary signal, which is typically required in QKD implementations. In our case, this signal allows to frequency-lock Alice and Bob's lasers. This scheme enhances the spatial sharing capabilities of the MCF, which is more attractive than dedicating a single fiber for this purpose, and an interesting alternative to wavelength-sharing over a SMF. Finally, our results push forward the deployment of CV-QKD systems over metropolitan networks.

## 2. Description of the experimental setup

Our experimental setup is shown in Fig. 1. In Inset (A), we schematically depict the transmitted signal from Alice to Bob, where reference pulses are interleaved between the pulses containing the quantum states. The reference pulses were used to recover the phase of the quantum states by allowing not only the estimation of the phase difference between the laser source and the LO, but also the phase variations caused by the channel [27, 28]. The pulses have a width of 4 ns with a pulse rate ($R$) of 31.25 Mpulses/s.

Alice's source was a 10 kHz linewidth continuous-wave (CW) external cavity laser (ECL) with tunable single-mode operation in the whole C band. This laser was frequency-locked with Bob's LO at 1550 nm (details below). The output of the ECL was fed into four electro-optic modulators connected in series. The first two modulators, corresponding to a phase modulator (PM) and an amplitude modulator (AM1), were used to provide the coherent states of light according to the GMCS protocol in which both X and P quadratures are independent and modulated following a zero-centered Gaussian random distribution [23]. The second amplitude modulator (AM2) was used to increase the extinction ratio of the light pulses while the third amplitude modulator (AM3) was used to control the ratio between the intensity of the reference and quantum pulses ($\rho$) by adjusting the bias set-point. A $\rho$ value of 300 was established to ensure a good accuracy in the phase recovery process [29]. The electro-optic modulators were driven with a field programmable gate array (FPGA) electronic board with a 1 GS/s digital-to-analog converter unit. Two independent sequences of $2^{11}$ pseudo-random values were used to generate the quantum states. The first one, related to the phase modulation, in which the values were uniformly distributed between 0 and $2V_{\pi-PM}$, with $V_{\pi-PM}$ being the half-wave voltage of the PM. The second sequence is related to the amplitude modulation, in which the values followed a Rayleigh distribution between 0 and the half-wave voltage of the amplitude modulator AM1 ($V_{\pi-AM1}$). For the reference pulses, their amplitude and phase always remained constant. After modulator AM3, a 50:50 fiber beam splitter (BS1) and a PIN photodiode (PD1) were employed to automatically optimize the bias set-point of the modulators by means of the FPGA. An electronic variable optical attenuator (VOA) was used to adjust Alice's modulation variance, defined as $V_{mod} = 2\langle n \rangle$, where $\langle n \rangle$ is the mean photon number measured at Alice's output. To determine $\langle n \rangle$, a 99:1 fiber beam splitter (BS2) sending 99% of the light to an optical power meter (PoM) was used. The light from the 1% output of BS2 was further divided by a $1 \times 8$ fiber beam splitter to simulate signals coming from six different QKD transmitters. The signals were connected to a fan-in device (Optoscribe), used to multiplex the signals into a 15 km 7-core MCF fiber spool developed by Corning Incorporated. After transmission, the signals were demultiplexed using a fan-out device (Optoscribe). Inset (B) in Fig. 1 shows the laboratory setup used for the 15 km 7-core MCF spool under test. Including the fan-in and -out devices, the average insertion loss per core was 6.3 dB, the forward(backward) crosstalk was lower than 50(74) dB and the backscattering loss was lower than 33 dB, see Fig. 2.

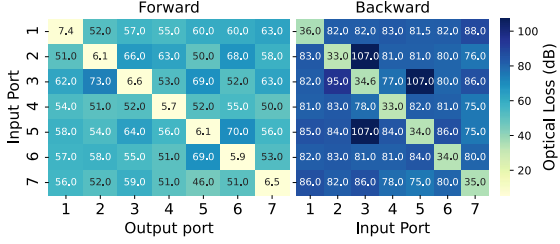

**Fig. 2.** 15 km 7-core MCF characterization: optical loss including fan-in and out devices in forward and backward directions.

Bob, which was connected to each core of the MCF under measurement, consisted of the LO laser, another 10 kHz linewidth CW ECL biased to emit 48 mW of optical power. Its output is split in two using a 99:1 fiber beam splitter (BS3), with 1% transmitted power to Alice through core 1 (C#1) in order to lock the frequency of Alice's laser to the LO. C#1 was chosen because it exhibited the highest insertion loss (7.4 dB, see Fig. 2). At Alice's side, the frequency difference between both lasers was minimized by applying a voltage to a piezoelectric element present in the cavity of Alice's laser. This minimization process was performed in real time by the FPGA through the measurement of the signal resulting from the interference between Alice's laser and the LO. For this purpose, a PIN photodiode (PD2) and two 50:50 fiber beam splitters (BS4 and BS5) were used as it can be observed in Fig. 1.

The other output of BS3 in Bob's side was connected to a 90° optical hybrid (90° OH), which was then followed by two balanced photodetectors (BPDs). Both X and P quadratures were simultaneously digitized with a 1 GS/s real-time oscilloscope (RTO). To calibrate the shot noise variance, an optical switch (SW 1) was used to block the incoming light from Alice. The electronic noise variance was measured by switching off a second switch (SW 2) located at LO input port of the Bob's OH, shown in Fig. 1 [28, 30]. Finally, the RTO bandwidth was fixed to 200 MHz, obtaining an average clearance value over the bandwidth of 16.7 dB. The clearance is defined as the power ratio between the shot noise and electronic noise.

A manual polarization controller was used at Alice and Bob's inputs to maximize overlap of the signals. To prevent Trojan horse attacks, an optical isolator was placed at Alice and Bob's output [31].

The post-processing required to characterize the proposed system was performed offline. It included down-sampling, quantum state phase recovery, pattern synchronization, and estimation of different parameters such as excess noise, channel transmittance and secret key rate. The down-sampling was based on detecting the power peaks of received signals. The phase shift of the $i^{th}$ quantum state ($\Delta\emptyset_{QS_i}$) was calculated by a linear interpolation between the phase of the $i^{th}$ and $i^{th} + 1$ references ($\emptyset_{R_i}$ and $\emptyset_{R_{i+1}}$) as indicated in Eq. (1) [27, 28]. $\emptyset_{R_i}$ is given by $arctg(X_{R_i}/P_{R_i})$ with $X_{R_i}$ and $P_{R_i}$ being the received quadratures of the $i^{th}$ reference. Regarding pattern synchronization, the time offset was removed by performing the cross-correlation between the transmitted and received quantum states.

$$\Delta\emptyset_{QS_i} = (\emptyset_{R_i} + \emptyset_{R_{i+1}})/2 \quad (1)$$

The excess noise at Bob ($\varepsilon$), measured in shot-noise unit (SNU), was estimated using Eq. (2) [28, 30], where $v_{elec}$ is the electronic noise variance, 1 stands for the shot noise variance and $V_{B|A}$ is the conditional variance given by Eq. (3) [28, 30]. In Eq. (3), $\eta$ is the detection efficiency, $q_A = \{X_{Ai}, P_{Ai}\}$ and $q_B = \{X_{Bi}, P_{Bi}\}$ are Alice and Bob's quadrature data, respectively, and $T$ is the channel transmittance. The channel loss was supposed to be controlled by an eavesdropper, therefore, $T$ can be estimated as indicated in Eq. (4) [28, 30], being $\langle q_A q_B \rangle$ the inner product between Alice and Bob's data.

$$\varepsilon = 2(V_{B|A} - 1 - v_{elec}) \quad (2)$$

$$V_{B|A} = var\left(\sqrt{\frac{\eta T}{2}} q_A - q_B\right) \quad (3)$$

$$T = \frac{2}{\eta}\left(\frac{\langle q_A q_B \rangle}{V_{mod}}\right)^2 \quad (4)$$

Finally, the secret key rate ($SKR$) in bits per second was estimated assuming the asymptotic limit and the reverse reconciliation as presented in Eq. (5) [28, 30], where $\beta$ [32], $I_{AB}$, $X_{BE}$ and $R_{eff}$ are the reconciliation efficiency, the mutual information between Alice and Bob, the Holevo bound and the effective quantum pulse rate, respectively. In this case, $I_{AB}$ can be determined from experimental parameters previously defined as it is expressed in Eq. (6) [28, 30].

$$SKR = (\beta I_{AB} - X_{BE})R_{eff} \quad (5)$$

$$I_{AB} = \log_2\left(1 + \frac{\eta T V_{mod}}{2 + \varepsilon + 2\nu_{elec}}\right) \quad (6)$$

## 3. Experimental results

Figs. (3)-(5) show the experimental results for the proposed CV-QKD system using the 15 km 7-core MCF with C#2 under test (see Fig. (1)). Transmission parameters are summarized in Table 1. Fig. 3 shows the phase-space density of received signal before and after the phase recovery process. In Fig. 3(b), the outer symbols correspond to the references and the inner ones are the quantum symbols corresponding to the received quantum states.

**Table 1.** Summary of the transmission parameter.

| Parameter | Value |
|---|---|
| Alice's modulation variance, $V_{mod}$ | 1.764 SNU |
| Electronic noise variance, $\nu_{elec}$ | 21 mSNU |
| Reconciliation efficiency, $\beta$ [32] | 0.95 |
| Detection efficiency, $\eta$ | 0.18 |
| Intensity ratio between reference and quantum symbols, $\rho$ | 300 |
| Effective quantum pulse rate, $R_{eff}$ | 15.625 Mpulses/s |

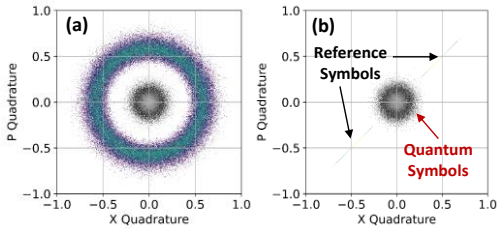

**Fig. 3**. Phase-space density in arbitrary units before (a) and after (b) phase recovery using reference symbols when C#2 of the 15 km 7-core MCF is under test.

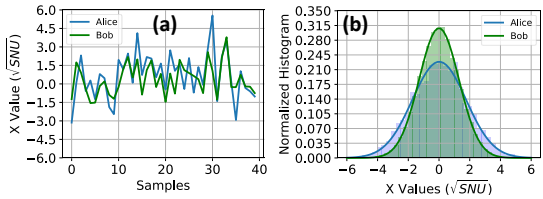

**Fig. 4.** (a) Comparison of the first 40 symbols of Alice and Bob's data for X quadrature when C#2 of the 15 km 7-core MCF is under test. (b) Histogram of the X quadrature data for Alice and Bob.

Fig. 4(a) shows 40 symbols of X quadrature for Alice and Bob. Bob's data corresponds to the detected signal after the phase recovery process.

High correlation can be seen. As it can be observed in Fig. 4(b), Alice and Bob's quadratures follow zero-centered Gaussian distributions. The variance of Bob's quadrature distribution is $V_B = (\eta T V_{mod} + \varepsilon)/2 + \nu_{elec} + 1$ [28, 30].

Fig. 5 shows 30 consecutive measurements taken over 90 minutes of the excess noise $\varepsilon$ (Fig. 5(a)), the channel transmittance $T$ and the secret key rate $SKR$ (Fig. 5(b)). In each measurement, a block size of $10^6$ symbols (coherent states) is considered. The $\varepsilon$, $T$ and $SKR$ values are calculated for each block of symbols independently. The shot-noise and electronic noise variances are calibrated before each measurement. The excess noise $\varepsilon$, defined in Eq. (6), has a mean value of 12 mSNU and always remains below the threshold in which the secret key rate is null, as exhibited in Fig. 5(a). The measured $\varepsilon$ threshold value, which depends on the experimental parameters, is 45 mSNU. The excess noise includes the phase estimation error due to time-delay between reference and quantum symbols and the noise owing to the reference symbols. The transmittance $T$, which is given by Eq. (4), presents a mean value of 0.673. In Fig. 5(b), it can be noticed that the value $T$ is quite stable over the measurement time with a relative standard deviation of 0.5%. The $SKR$ is determined considering the values obtained for $\varepsilon$ and $T$ according to Eq. (5). A mean $SKR$ of 0.372 Mb/s is achieved during the measurement time with minimal and maximal values of 0.254 Mb/s and 0.533 Mb/s, respectively, as shown in Fig. 5(b).

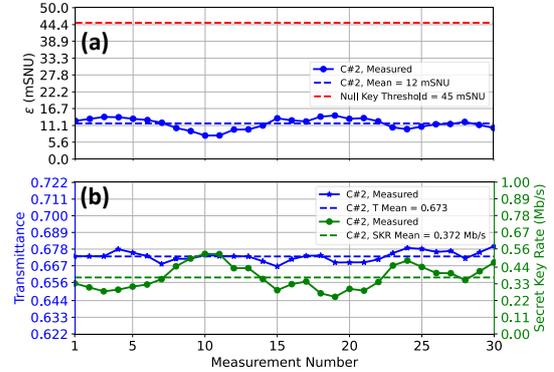

**Fig. 5.** Obtained results for 30 measurements during 90 minutes when C#2 of the 15 km 7-core MCF is under test: (a) excess noise ($\varepsilon$) and (b) channel transmittance ($T$) and secret key rate ($SKR$). Each measurement corresponds to a block size of $10^6$ symbols or coherent states.

Table 2 shows the CV-QKD system characterization using the 15 km 7-core MCF in terms of $\varepsilon$, $T$ and $SKR$ parametrs for each core. There, obtained values per core are the average of 30 measurements with a block size of $10^6$ symbols per measurement. From Table 2, it can be observed that obtained results are pretty similar across all cores, achieving a potential aggregated $SKR$ value of 2.3 Mb/s. This potential aggregated $SKR$ is defined as the sum of all the obtained *SKR*s shown in Table 2. Since the light coming from the QKD transmitter is divided into 6 cores, each of them measured individually at Bob, our experiment can be considered as a simulation of a scenario where 6 transmitters and 6 receivers are used.

**Table 2.** Summary of characterization of the proposed CV-QKD system using the 15 km 7-core MCF in terms of excess noise ($\varepsilon$), channel transmittance ($T$) and secret key rate ($SKR$) for each core. Obtained values are an average of 30 measurements with a block size of $10^6$ symbols per measurement.

| C# | $\varepsilon$ (mSNU) | $T$ | $SKR$ (Mb/s) |
|---|---|---|---|
| 2 | 11.8 | 0.673 | 0.372 |
| 3 | 10.2 | 0.650 | 0.378 |
| 4 | 11.6 | 0.678 | 0.378 |
| 5 | 12.4 | 0.692 | 0.389 |
| 6 | 14.7 | 0.667 | 0.371 |
| 7 | 12.3 | 0.645 | 0.375 |

## 4. Conclusions

In conclusion, we have demonstrated CV-QKD across a 15 km MCF where one of the cores is used to transmit an auxiliary signal to frequency-lock the two lasers. We have also demonstrated the capacity of the MCF to boost the secret key rate by parallelizing CV-QKD across multiple cores. In particular, a potential aggregated *SKR* value of 2.3 Mb/s has been achieved. Finally, our results push forward the deployment of CV-QKD systems over metropolitan regions taking advantage of next-generation telecom optical fibers.

## Funding

Horizon 2020 Framework Programme (820466), H2020 Marie Skłodowska-Curie Actions (713729), Ceniit Linköping University, the Swedish Research Council (2017-04470) and QuantERA grant SECRET (2019-00392).